\documentclass{elsart}

\usepackage{graphicx}

\begin{document}

\begin{frontmatter}

\title{Hypernuclear Physics for Neutron Stars}

\author{J\"urgen Schaffner-Bielich}

\address{Institut f\"ur Theoretische Physik/Astrophysik, J. W. Goethe
  Universit\"at\\ D-60438 Frankfurt am Main, Germany}

\ead{schaffner@astro.uni-frankfurt.de}

\begin{abstract}
  The role of hypernuclear physics for the physics of neutron stars is
  delineated. Hypernuclear potentials in dense matter control the
  hyperon composition of dense neutron star matter. The three-body
  interactions of nucleons and hyperons determine the stiffness of the
  neutron star equation of state and thereby the maximum neutron star
  mass. Two-body hyperon-nucleon and hyperon-hyperon interactions give
  rise to hyperon pairing which exponentially suppresses cooling of
  neutron stars via the direct hyperon URCA processes. Non-mesonic weak
  reactions with hyperons in dense neutron star matter govern the
  gravitational wave emissions due to the r-mode instability of rotating
  neutron stars. 
\end{abstract}


\end{frontmatter}


\section{Introduction}


Neutron stars are born in spectacular core collapse supernova
explosions. These compact, massive objects have typical radii of about
10 km and masses of $(1-2)M_\odot$. Matter in the core of neutron stars is
compressed to extreme densities, several times normal nuclear matter
density, i.e.\ $\rho\gg \rho_0 = 3\cdot 10^{14}$~g/cm$^3$. The relation of
supernova explosions being the birthplace of neutron stars is
exemplified by the historic supernova remnant of AD 1054, the crab
nebula, and the crab pulsar, a rotation-powered neutron star,
sitting in its center.

More than 1700 pulsars are recorded in the publicly available pulsar
data base at the Australian National Telescope Facility
\cite{Manchester:2004bp}. The number of discovered pulsars is
continuously growing with the ongoing pulsar surveys at radio
telescopes worldwide (Arecibo, Green Bank Telescope, Parkes Multibeam).
The best determined mass is still the one of the Hulse-Taylor pulsar
with $M=(1.4411\pm0.00035)M_\odot$ \cite{Weisberg:2004hi}, the fastest
rotating one is the pulsar PSR J1748-2446ad with 716 revolutions per
second \cite{Hessels:2006ze}. 

Recently, indications of an extremely massive neutron star have been
published \cite{Freire:2007jd,Freire:2007ux}. If the measured periastron
advance is due to effects from general relativity only, the pulsar PSR
J1748-2021B has a mass of $M\geq (2.74 \pm 0.21) M_\odot$ and is more
massive than $M\geq 2.0 M_\odot$ with a confidence level of 99\%. Note,
that it can not be excluded at present that the system measured actually
contains two neutron stars, not a single one. Redshifted spectral lines
have been claimed to be extracted from the analysis of x-ray
bursts from EXO 0748--676 \cite{Cottam:2002cu}, which give a constraint
on the mass-radius ratio of the compact star. A recent analysis 
comes to the conclusion that the compact star mass is $M\geq
(2.10\pm 0.28) M_\odot$ with a radius of $R\geq (13.8 \pm 1.8)$~km
\cite{Ozel:2006km} claiming that 'unconfined quarks do not exist at the
center of neutron stars'! However, this conclusion was put into
perspective in a follow-up reply \cite{Alford:2006vz} which demonstrated
that those limits rule out soft equations of state, but not quark
stars or hybrid stars. The interactions between quarks can be quite strong
so that the presence of quark matter in the core stabilises the
compact star. On the other hand, the mass limit provides indeed a strong
constraint for hyperons in dense neutron star matter. Hyperons are
likely to appear at moderate densities, which will substantially
decrease the maximum mass. This conclusions is guided by hypernuclear
data and present model calculations. If such massive neutron stars are
confirmed in the future, say with masses above $2M_\odot$, then it seems
that our present understanding of hypernuclear physics of compact stars
will be in conflict with the pulsar data, as we will outline in more
detail below.

Constraints on the mass and radius of neutron stars can be derived by
observations in the optical as well as in the x-ray band, a booming
field of exploration since the launch of the x-ray satellites Chandra
and XMM-Newton in 1999. The best studied isolated neutron star is RXJ
1856.35-3754, the closest one known. A two-component blackbody fit to
the combined optical and x-ray spectra results in a low soft
temperature, so as not to be in contradiction with the observed x-ray
flux. This low temperature implies a rather large radiation radius, the
radius observed at infinity, $R_\infty = R/\sqrt{1-2GM/R}$, so that the optical
flux comes out right. A conservative lower limit was given in
\cite{Trumper:2003we} and confirmed in a detailed modelling of the
neutron star atmosphere \cite{Ho:2006uk,Ho:2007aw} as being $R_\infty
\approx 17$~km for an assumed distance of $d=140$~pc. With the derived
gravitational redshift of $z_g\approx 0.22$, the true radius of the
neutron star would be about $R\approx 14$~km with a corresponding mass
of $M\approx 1.55M_\odot$.  The large radiation radius implies generally
a large neutron star radius which could only be explained with a stiff
nuclear equation of state.  The biggest uncertainty, besides the
systematical one, is the distance to the neutron star.

The spectra of the neutron star X7 in the Globular Cluster 47 was fitted
in Ref.~\cite{Rybicki:2005id} with an improved hydrogen atmosphere
model.  The radius of the neutron star was estimated to be $R_{\rm
  ns}=(14.5^{+1.8}_{-1.6})$~km, which is a little bit larger than for
non-relativistic nuclear models but right in the range of standard
relativistic mean-field models, see e.g.~\cite{Lattimer:2006xb}.  The
authors of Ref.~\cite{Rybicki:2005id} state, that for a radius of 10 km
the mass should be in the range $M_{\rm ns}=(2.20^{+0.03}_{-0.16})
M_\odot$.  For a radius of 14~km, however, any mass between 0.5 and
$2.3M_\odot$ is allowed by the fit.  On the other hand, atmosphere fits
to the spectra of M13 lead to rather small radii, a radius of only
$R=9.77^{+0.09}_{-0.29}$~km was derived in Ref.~\cite{Webb:2007tc}.  The
allowed ranges in the mass-radius diagram for the fit to the spectra of
M13 and X7 are nearly mutually exclusive on the 99\% confidence level.
However, one should keep in mind, that the whole mass-radius curve for
neutron stars just has to reach somewhere those two regions. In fact,
many of the mass-radius curves shown in \cite{Webb:2007tc} pass the two
constraints from the fits to the spectra of M13 and X7, except for the
curves of the most stiffest models, in particular for the relativistic
field theoretical models without hyperons.

Another way of probing neutron star matter properties is by cooling
observations of supernova remnants, see e.g.\
\cite{Kaplan:2004tm,Kaplan:2006mb}. The observational limits points
towards fast cooling processes in the interior of neutron stars, i.e.\
direct URCA reactions. Standard conventional cooling curves are too
high, so that either a large nuclear asymmetry energy or strange exotic
particles are needed to generate efficient and fast cooling (see
\cite{Yakovlev:2004iq} for a theoretical review).

The basic structure of the low-density region of neutron stars is fairly
well-known. The outer crust consists of a lattice of nuclei with free
electrons and is a few 100 meters thick. The sequence of nuclei is
controlled by their binding energies and follows mainly along the
neutron magic numbers 50 and 82 (for a most recent investigation of the
outer crust see \cite{Ruster:2005fm}). Similar features will be
discussed in the context of hypernuclei below. The inner crust starts at
the neutron drip density at $n\approx 4\cdot 10^{11}$~g/cm$^3$ and
consists of a lattice of nuclei with free neutrons and electrons. The
core starts at the end of the inner crust which occurs around half times
normal nuclear matter density. In this core region, hyperons can
populate the interior of neutron stars. The implications of the presence
of hyperons for the properties of neutron star will be outlined in this
review, which is an update and an extension of a preliminary version of
Ref.~\cite{SchaffnerBielich:2007tj}.


\section{Hyperons in Neutron Stars!}


The term neutron star implies that the main component of neutron star
matter are just neutrons. However, this picture changes drastically for
matter at extremely high densities, i.e.\ in the core of neutron stars.
Simple arguments for the presence of other more exotic species besides
nucleons, electrons and muons can be given in terms of a free gas of
hadrons and leptons. Matter in $\beta$-equilibrium but with no
interactions starts to populate $\Sigma^-$ hyperons already at 4$n_0$,
where $n_0$ is the normal nuclear matter density, the lighter $\Lambda$
hyperons appear at $8n_0$ \cite{Ambart60}. Inclusion of nuclear forces
generically reduces these critical densities substantially, so that
hyperons appear already around $2n_0$ (see e.g.\ \cite{Sahakian:1963}
and references therein for the very first investigations of this kind).

That interactions are essential for the description of neutron star
properties is evident from the fact that the corresponding equation of
state of a free gas results in a maximum mass of only $M_{\rm max}
\approx 0.7 M_\odot$ \cite{OV39} which is by more than a
factor two smaller than the presently most precisely known pulsar mass
of $1.44 M_\odot$ for the pulsar PSR 1913+16. Hence, effects from strong
interactions are crucial in describing neutron stars raising the maximum
mass from 0.7 to two or more solar masses \cite{Cameron59}.
Note, that this is in contrast to white dwarfs which are basically
stabilised by the Fermi pressure of the free electron gas only.

As hyperons are likely to be present in addition to nucleons, one has to
consider the interactions between all stable baryons. Besides the
nuclear force, there is some knowledge from hypernuclear physics about
the interactions between hyperons and nucleons and scarcely between
hyperons themselves. The $\Lambda N$ interactions is very well studied,
the potential depth of $\Lambda$ hyperons is $U_\Lambda = -30$~MeV at
$n=n_0$ (see e.g.\ \cite{Millener88}), so that bound $\Lambda$
hypernuclear states exists. The situation is different for $\Sigma$
hyperons. The only bound $\Sigma$ hypernucleus known so far,
$^4_\Sigma$He, is bound by isospin forces \cite{Hayano89,Nagae98}. A
detailed scan for $\Sigma$ hypernuclear states turned out to give
negative results \cite{Bart99}. The study of $\Sigma^-$ atoms shows
strong evidence for a sizable repulsive potential in the nuclear core,
i.e. at $n=n_0$ \cite{Batty:1994yx,Batty:1994sw,Mares:1995bm}. A recent
review on hadronic atoms can be found in \cite{Friedman:2007qx} which
confirms the repulsive nature of the nuclear $\Sigma^-$ potential within
a new geometric analysis of the $\Sigma^-$ atomic data. On the other
hand, the $\Xi$ nucleon interactions seems to be attractive, several
$\Xi$ hypernuclear states are reported in the literature \cite{Dover83}.
More recently, quasi-free production of $\Xi$'s reveal an attractive
potential of $U_\Xi=-18$~MeV \cite{Fukuda98,Khaustov2000} (with
relativistic corrections, see \cite{SchaffnerBielich:2000wj}).  Last but
not least, the hyperon-hyperon (YY) interaction is not really well
known, there are just a few double $\Lambda$ hypernuclear events (for a
recent review see \cite{Gal:2003ze}). The interaction between other
pairs of hyperons as $\Lambda\Xi$ or $\Xi\Xi$ is not known at all
experimentally. However, the hyperon potentials are essential for the
determination of the composition of neutron star matter so basic
hypernuclear data can provide substantial input for the modelling of
neutron star matter.

Important for the stability of neutron stars is the short-range
repulsion of the baryon-baryon interaction. Fits with nonrelativistic
potentials to $\Lambda$ hypernuclear data show effects from three-body
interactions for the $\Lambda N$ interaction \cite{Millener88}. The
density dependence of the Schr\"odinger equivalent potential is
compatible with the many-body mean-field potential of relativistic
field-theoretical approaches and demonstrates that the hyperon potential
turns repulsive above $2n_0$ \cite{Schaffner:1996kv}. The absence of
these higher-order terms in density is likely to generate too soft an
equation of state, so that the maximum mass of neutron stars falls below
the mass limit of $1.44M_\odot$. Arguably, this might be the reason that
modern many-body calculations of neutron star matter with
nucleons and hyperons result in too low neutron star maximum masses
\cite{Baldo:1999rq,Vidana:2000ew,Nishizaki:2002ih,Schulze:2006vw}.
Additional repulsion between hyperons and nucleons is needed. The
hyperon three-body force has not received too much attention
but is known for quite some time to be repulsive in nature for $\Lambda
NN$ \cite{Gal:1967} leading to the needed additional stability for
neutron stars.

\begin{figure}
\vspace*{-1cm}
\centering{\includegraphics[width=0.9\textwidth]{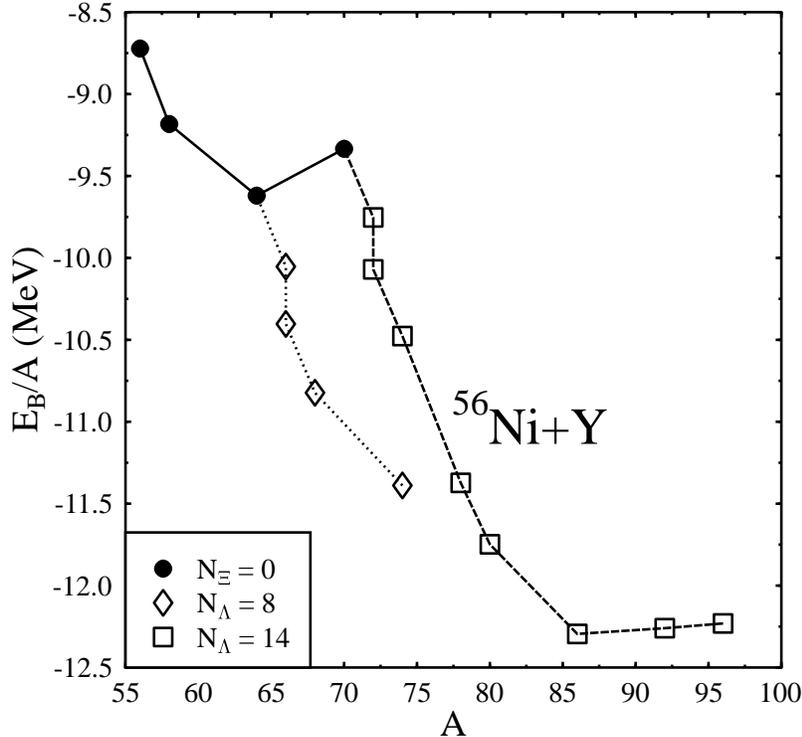}}
\vspace*{-1cm}
\caption{The binding energy of strange hadronic matter for a nucleonic
   core of $^{56}$Ni with added $\Lambda$ and $\Xi$ hyperons as a
   function of baryon number $A$ (taken from \cite{Scha93}).}
\label{fig:shm}
\end{figure}

The appearance of hyperons in dense neutron star matter can be also
elucidated by looking at finite systems of nucleons and hyperons, so
called strange hadronic matter
\cite{Scha92,Scha93,Scha94,SchaffnerBielich:2000wj}. Let us consider an
arbitrary number of nucleons and hyperons forming one big
multi-hypernucleus. The system is stable against strong interactions, if
reactions as $\Lambda + \Lambda \leftrightarrow \Xi + N$ and $\Sigma + N
\rightarrow \Lambda + N$ are Pauli-blocked. The first reaction releases
an energy of $Q\approx 25$~MeV, the second one $Q\approx 80$~MeV so that
$\Sigma$ hyperons can be hardly stabilised in hypernuclear systems. A
similar feature will be present for neutron star matter, where it is
indeed also likely that $\Sigma$ hyperons do not appear (although the
main reason is due to the repulsive potential for $\Sigma$ hyperons).
One can construct stable systems of nucleons and hyperons by adding
successively $\Lambda$ hyperons until $\Xi$ hyperons can be populated as
the filled $\Lambda$ hypernuclear levels prevent the strong reactions by
Pauli-blocking. Fig.~\ref{fig:shm} shows the binding energy of such
Pauli-blocked systems for a nucleonic core of $^{56}$Ni versus the
baryon number. When the p-shell of the $\Lambda$ hypernuclear level is
filled up, $\Xi$ hyperons can be added in the s-shell without loosing
stability. The addition of hyperons leads to an overall
increase in the binding energy as the hyperons populate deep lying s--
and p-- states in a separate quantum well.  The nuclear binding energy
with $\Lambda$s and $\Xi$s reaches up to $E/A = -12$~MeV (here a weak YY
interaction is assumed)! In terms of the binding energy, it is
energetically favoured to add hyperons to the system. A similar effect
occurs for dense matter in $\beta$-equilibrium: here beyond some
critical density, the filling of low-lying (with low Fermi momenta)
hyperon states in a newly opened quantum well becomes preferred compared
to adding more nucleons at large Fermi momenta. Hyperons appear in dense
matter when their in-medium energy $\omega(Y)$ equals their chemical potential
$\mu(Y) = \omega(Y) = m_Y + U_Y(n)$. Hyperons are then Pauli-blocked and
can not decay as all levels are filled up for its possible decay
products. In the case of neutron star matter, strange hadronic matter
becomes now even stable to weak interactions!

In modern nuclear models, which are fitted to nuclear and hypernuclear
data, hyperons appear in neutron star matter at $n\approx 2n_0$ in
relativistic mean-field (RMF) models
\cite{Glendenning:1984jr,Schaffner:1995th,Knorren:1995ds}, in a
nonrelativistic potential model \cite{Balberg:1997yw}, in the quark-meson
coupling model \cite{Pal:1999sq}, in relativistic Hartree--Fock models
\cite{Huber:1997mg}, in Brueckner--Hartree--Fock calculations
\cite{Baldo:1998hd,Baldo:1999rq,Vidana:2000ew,Schulze:2006vw}, in chiral
effective Lagrangians \cite{Hanauske:1999ga}, in the density-dependent hadron
field theory \cite{Hofmann:2000mc}, and in G-matrix calculations
\cite{Nishizaki:2002ih}.  It is remarkable that one of the very first
calculations came to a similar conclusion \cite{Sahakian:1963}. Hence,
neutron stars are indeed giant hypernuclei \cite{Glendenning:1984jr}!

\begin{figure}
\centering{\includegraphics[width=0.9\textwidth]{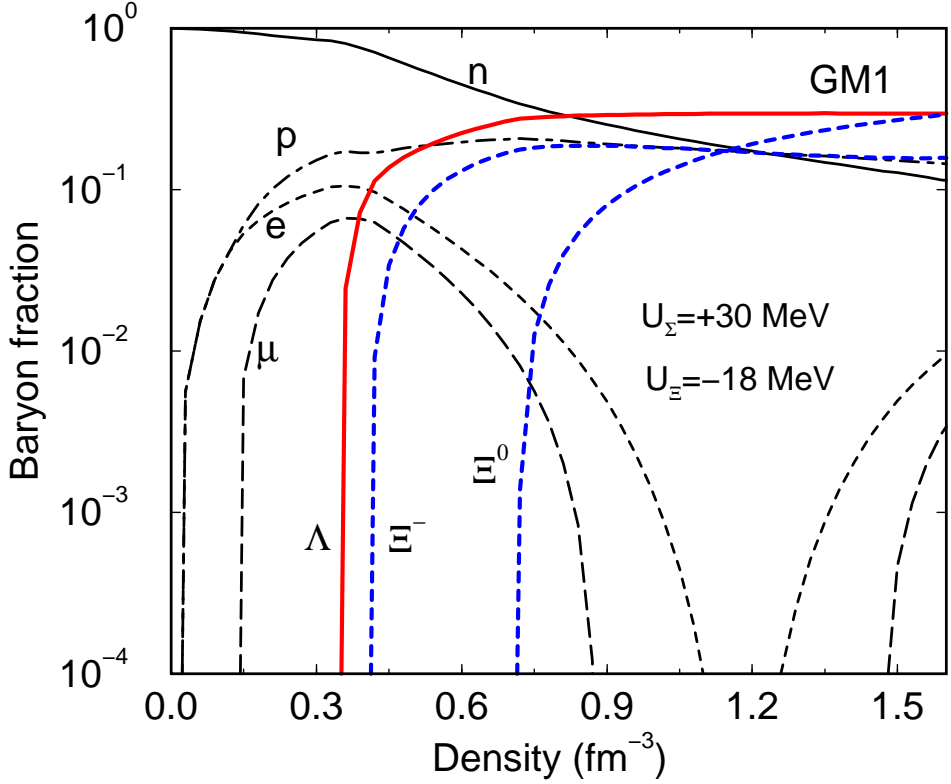}}
\caption{The fraction of baryons and leptons in neutron star matter for
  a RMF calculation using set GM1 with weak hyperon-hyperon interactions
  (see \cite{Schaffner:1995th}).}
\label{fig:comp}
\end{figure}

The composition of neutron star matter depends sensitively on the
assumed hypernuclear potentials. The $\Sigma^-$ hyperon appears in dense
matter usually together with the $\Lambda$ at about $2n_0$, in some
cases even slightly before the $\Lambda$ due to its negative charge, if
an attractive potential of $U_\Sigma=-30$~MeV similar to the $\Lambda$
is chosen. However, for a repulsive potential the $\Sigma^-$ as well as
the other $\Sigma$ hyperons will not be present in neutron star matter
at all. Fig.~\ref{fig:comp} depicts the fraction of baryons and leptons
as a function of density for a relativistic mean-field calculation using
the parameter set GM1 \cite{Glendenning:1991es} assuming a repulsive
$\Sigma$ potential.  The $\Lambda$ is present at $2.3n_0$, the $\Xi^-$
hyperon at $2.7n_0$ (here the model with weak YY interaction is taken
from \cite{Schaffner:1995th}). Besides the $\Xi^0$ emerging at $4.7n_0$
no other hyperon is present up to $10n_0$, which is well beyond the
maximum density reached for this equation of state. It is clear that
hypernuclear data provides as an essential ingredient the hyperon
potential depth which controls the composition in the core of neutron
stars. The baryon and lepton population is highly sensitive to the
in-medium potential of hyperons which will turn out to be important for
the cooling of neutron stars.


\section{Hyperons and cooling of neutron stars}


Moderately aged neutron stars up to 1 million years after their
formation will dominantly cool by volume emission of neutrinos. Cooling
of photons from the surface will take over afterwards. The standard
reaction for cooling is the modified URCA processes $N + p + e^- \to N +
n + \nu_e$ and $N + n \to N + p + e^- + \bar\nu_e$ with a bystander
nucleon to conserve energy and momentum. The modified URCA process is
slow and leaves the neutron star quite warm until the photon cooling
epoch. Much faster reactions are the direct URCA processes as $p + e^-
\to n + \nu_e$ and $n \to p + e^- + \bar\nu_e $. However, these reactions
can only proceed if the Fermi momenta fulfil the condition $p_F^p +
p_F^e \geq p_F^n$. Charge neutrality implies $n_p = n_e$ or $p_F^p =
p_F^e$, so that $2p_F^p \geq p_F^n$. Hence, the proton fraction has to
exceed $n_p/n \geq 1/9\approx 11\%$ for the direct nucleon URCA process
to start. Relativistic calculations usually reach this value quite
easily \cite{Lattimer:1991ib}. From Fig.~\ref{fig:comp} one can read off
the critical density for the direct nucleon URCA process to be $1.5n_0$.
Nonrelativistic calculations do not get that large proton fraction, as
the asymmetry energy does not have the same strong density dependence as in
relativistic models. In addition, nucleons are pairing strongly, so that
energy is needed to break them up (recent reviews on cooling of neutron
stars can be found in \cite{Yakovlev:2004iq,Sedrakian:2006mq}).

On the other hand, hyperons can help substantially to cool a neutron
star via the hyperon direct URCA processes as $\Lambda \to p + e^- +
\bar\nu_e$ or $\Sigma^- \to \Lambda + e^- + \bar\nu_e$. Remarkably, the
hyperon direct URCA processes happen immediately when hyperons are present
and can also occur if there is no direct URCA process for nucleons
allowed \cite{Prakash:1992}! There is no minimum fraction of hyperons needed,
as there is no additional constraint from the charge neutrality
condition as for nucleons (in reality the presence of muons gives a
small critical fraction of a few per mille, see \cite{Prakash:1992}).
Hence, if nucleons are gapped the most important cooling mechanism
involves hyperons. 

For weak YY coupling or interaction strengths, there will be rapid
cooling due to the presence of hyperons mimicking some more exotic agent
as kaon condensation or quark matter in the core. The rapid cooling
process can start basically as soon as hyperons are part of the
composition of neutron star matter, which implies that there is some
critical neutron star mass for fast cooling.  Hyperon cooling is only
suppressed by hyperon pairing gaps which are presumably much smaller
than the ones for nucleons. The importance of hyperon superfluidity for
the hyperon direct URCA processes has already been pointed out in
Ref.~\cite{Haensel:1994}. Hence, a detailed modelling of the cooling of
neutron stars demands to have a knowledge not only on the composition,
which is fixed by the in-medium potential of hyperons, but also on the
YY interaction strength which determines the hyperon gap energy. There
exist a few studies on hyperon cooling in the literature (see
\cite{Schaab:1998zq,Page:2000wt,Vidana:2004rd,Zhou:2005hv,Takatsuka:2005bp}
and references therein). In the first hyperon cooling calculation with
hyperon pairing \cite{Schaab:1998zq}, hyperons are present in the core
for $M \geq 1.35 M_\odot$. An attractive $\Sigma$ nuclear potential was
adopted so that the $\Sigma^-$ appears even before the $\Lambda$. The
dominant cooling process involves the reaction $\Sigma^- \to \Lambda +
e^- + \bar\nu_e$. Two-body YY interactions were used as input to model
the hyperon pairing gaps and their emissivities. It was found that
hyperon gaps improve the thermal history and are more consistent with
x-ray observations of neutron stars. On the other hand, in a subsequent
study \cite{Page:2000wt} the $\Lambda$ hyperon appeared at a slightly
lower density than the $\Sigma^-$, so that there was a tiny density
range of unpaired $\Lambda$ hyperons present. These unpaired hyperons
resulted in even faster cooling for heavier stars via the hyperon direct
URCA. Also, new improved hyperon-nucleon interactions find very large
pairing gaps for the $\Sigma$ hyperons which would suppress the hyperon
direct URCA processes involving $\Sigma$ hyperons, see
e.g.~\cite{Vidana:2004rd}. However, note that cooling processes with
$\Sigma$ hyperons are likely to be not present anyway for a repulsive
$\Sigma$ nuclear potential as the $\Sigma$ hyperons would not be part of
the neutron star matter composition. The conclusion is, that indeed
two-body forces between hyperons and nucleons have an enormous impact on
the cooling history of neutron stars. Hence, hypernuclear physics serves
as a key ingredient not only for the composition of dense neutron star
matter but also for the cooling history of neutron stars.


\section{Hyperons and the maximum mass of neutron stars}


It is known for quite some time, that hyperons have a significant effect
on the global properties of compact stars. As new degree of freedom,
which can populate new Fermi levels, hyperons can lower the overall
Fermi energy and momentum of baryons and leptons. Thereby, the total pressure
of the system for a given energy density is considerably lowered, which
implies that the equation of state is substantially softened.

The first consistent implementation of the relativistic $\Lambda$ hyperon potential
depth in neutron star matter was performed by
Glendenning and Moszkowski \cite{Glendenning:1991es} using a
relativistic field theoretical approach. The other hyperon potentials
were fixed by assuming universal coupling strengths for all hyperons by
setting the $\Sigma$ and $\Xi$ hyperon coupling constants equal to the
one of the $\Lambda$. Hence, hypernuclear constraints for $\Sigma$ and
$\Xi$ hyperons were not taken into account, in particular the $\Sigma$
potential is as attractive as that of the $\Lambda$. Follow-up
calculations adopt SU(3) symmetry for the vector coupling constants and
specify the scalar coupling via the different hyperon potentials, see
\cite{Schaffner:1995th}. The neutron star sequence with nucleons and
leptons only reached a maximum mass of $M\approx 2.3M_\odot$. A
substantial decrease of the maximum mass occurred once hyperons were
taken into account, with parameters fixed by hypernuclear data. The
maximum mass for such ``giant hypernuclei'' turned out to be now around
$M\approx 1.7 M_\odot$.  Moreover, they demonstrated that the case of
noninteracting hyperons results in a too low maximum mass, i.e.\
$M<1.4M_\odot$! Clearly, strong (repulsive) interactions between
hyperons have to be implemented for a consistent description of pulsar
masses.

The issue of the softness of the nuclear equation of state and the
maximum mass of neutron stars has received considerable renewed interest
recently due to the analysis of heavy-ion data. The focus will be here
on the analysis of strange particle production in heavy-ion collisions,
in particular the subthreshold production of kaons measured by the KaoS
collaboration \cite{Sturm:2000dm,Forster:2007qk} at GSI, Darmstadt. The
ratio of the multiplicities per baryon of the produced kaons in C+C and
in Au+Au collisions turns out to be rather insensitive to the underlying
microphysical input for the transport simulation, as the kaon-nucleon
optical potential, cross sections, lifetime of resonances etc. The
analysis of the data at different bombarding energies with transport
simulations arrives at the conclusion that the nuclear equation of state
should be rather soft at densities around $2-3n_0$
\cite{Hartnack:1993bq,Fuchs:2000kp,Hartnack:2005tr,Fuchs:2007vt}. The
extracted compression modulus turns out to be around 200~MeV for a
simple Skyrme-type parameterisation of the nuclear equation of state.

However, as outlined above, most recent pulsar data points towards quite
large masses or large radii which can be only reconciled with a rather
stiff nuclear equation of state. There seems to be a conflict between
heavy-ion data and pulsar observations which can be resolved actually,
see Refs.~\cite{Sagert:2007nt,Sagert:2007kx}. First, transport models
use actually the Schr\"odinger equivalent potential as input not the
nuclear equation of state. Second, the nuclear density ranges probed are
different for the production of kaons and the maximum mass of neutron
stars. Typically, the maximum central density reached in the center of
neutron stars amounts to about $(5-6)n_0$, which depends on the assumed
nuclear model. These values could be much larger. However, one hardly
finds a calculation in the literature with substantially lower values
for the maximum central densities. As stated above, kaon production in
heavy-ion collisions is sensitive up to $2.5n_0$. Therefore, there is a gap
in the nuclear density regions probed. The stiffness of the hadronic
equation of state above $2.5n_0$ controls the value of the maximum mass
achievable for neutron stars. Interestingly, this is the density regime
where hyperons presumably appear and modify the neutron star matter
properties significantly. These lines of arguments have been
cross-checked in a more detailed investigation using Skyrme-type and
relativistic mean-field models \cite{Sagert:2007nt,Sagert:2007kx}. The
'soft nuclear equation of state' extracted from heavy-ion data is indeed
compatible with the recent pulsar mass measurements when only nucleons
and leptons are considered as the basic constituents in neutron star
matter. The inclusion of hyperons, however, causes an equation of state
which turns to be very soft at high densities with the constraint from
heavy-ion data. The maximum mass reached for several nuclear equations
of state analysed within the relativistic mean-field model is just
$M=1.53M_\odot$ \cite{Sagert:2007kx}. If a more massive neutron star is
confirmed, the role of hyperons in neutron stars in combination with the
constraint from heavy-ion data needs to be reinvestigated.

Again, hyperons play a decisive role in compact star physics. The
feature, that hyperons lower drastically the maximum mass of neutron
stars became even more pronounced with modern many-body approaches to
neutron star matter beyond the mean-field approximation.  In
relativistic Hartree-Fock calculations, the maximum mass of neutron
stars was computed to be $M_{\rm max} = (1.4-1.8) M_\odot$ depending
sensitively on the chosen hyperon coupling strength \cite{Huber:1998hm}.
In Brueckner-Hartree-Fock approaches using Nijmegen soft-core
hyperon-nucleon (YN) potentials maximum masses of $M_{\rm max} = 1.47
M_\odot$ have been derived for the nucleon-nucleon and YN interactions
only and $M_{\rm max} = 1.34 M_\odot$ when including the YY interactions
\cite{Vidana:2000ew}. In the same approach, three-body forces for
nucleons have been included but none for the hyperons so that a maximum
mass of only $M_{\rm max} = 1.26 M_\odot$ was attained
\cite{Baldo:1999rq}. The latter maximum masses are even below the mass
limit of the Hulse-Taylor pulsar of $1.44 M_\odot$ and the corresponding
hyperonic equations of state are clearly ruled out by pulsar data.
Obviously, some additional hyperon physics is missing in those cases.
Presumably, three-body forces for hyperons will solve this problem, as they
are repulsive and will raise the maximum mass (some crude investigations
in this directions can be found in \cite{Nishizaki:2002ih} supporting
this statement). Here, input is needed from hypernuclear physics, not
only for the hyperon three-body force but also for the momentum
dependence of the hyperon interactions, as dense matter probes momenta
of the order of several hundred MeVs. Contrary to the widely used
standard mean-field and nonrelativistic approaches, Brueckner-type
approaches adopt momentum-dependent potentials which have to be fixed by
YN scattering and hypernuclear data.

The YY interaction is another important ingredient for the description
of neutron star matter. In fact, it is even possible to generate a new
class of compact stars, hyperon stars, besides ordinary white dwarfs and
neutron stars, by a new stable solution of the
Tolman-Oppenheimer-Volkoff equation \cite{SchaffnerBielich:2002ki}. By
increasing the overall strength of the YY interactions (in particular
the unknown $\Xi\Xi$ interaction which can be probed in heavy-ion
collisions however, see \cite{SchaffnerBielich:1999sy}), a first order
phase transition appears from neutron matter to hyperon-rich matter. A
mixed phase is present for a wide range of densities $n_{\rm
  mix}=(2.5-6.5)n_0$. Interestingly, all hyperons ($\Lambda$, $\Xi^0$,
$\Xi^-$) appear at the start of the mixed phase, as the bubbles of the
new hyperon phase are charged and have a larger density than the
surrounding normal neutron matter (note that for a Gibbs construction
the chemical potentials must be equal in phase equilibrium, not the
densities). The strong first order phase transition due to hyperons has
a strong impact on the mass-radius relation for compact stars.  A new
stable solution in the mass--radius diagram appears, as the curve
reaches a second maximum for the mass for small radii. Those hyperon
stars are generated via attractive YY interactions (mainly $\Xi\Xi$
interactions). We note that a weak $\Lambda\Lambda$ does not rule out a
strong $\Xi\Xi$ interaction nor the possible existence of hyperon stars.
Within the Nijmegen soft core model NSC97, the hyperon-hyperon
interactions are highly attractive in certain channels
\cite{Stoks:1999bz}. Even a bound $\Xi\Xi$ state was found. However,
other bound states also appear in the NSC97 model which are now
considered to be fictitious. The new Nijmegen potential ESC04
\cite{Rijken:2006ep,Rijken:2006kg} has not been extended to the $S=-4$
sector so far, unfortunately. In a recent SU(6) quark model calculation,
which derived the baryon potentials for the full baryon octet, no bound
$\Xi^0\Xi^0$ state has been found \cite{Fujiwara:2006yh}.

The two different solutions for hyperon-rich matter behave like neutron
star twins: they have similar maximum masses, $M_{\rm hyp} \sim M_n$,
but different radii $R_{\rm hyp}<R_n$. In addition, selfbound compact
stars for strong YY attraction with $R=7-8$ km are also possible, but
demand that strange hadronic matter is absolutely stable so that
ordinary neutron stars are completely converted to hyperon stars.  Such
neutron star twin solutions have been also found for a strong first
order phase transition to quark matter
\cite{Glendenning:1998ag,Schertler:2000xq,Fraga:2001id}.  In fact, any
strong first order phase transition can produce a so-called third family
of compact stars. Signals for such a strong phase transition can be
detected by direct mass and radius measurements, or by the collapse of a
neutron star to the third family via measurements of gravitational
waves, $\gamma$-rays, and neutrinos.

In passing, I note that strange multiquark states can also exist in
neutron stars, as the H-dibaryon \cite{Glendenning:1998wp} or strange pentaquarks
\cite{Sagert:2006na}. Pentaquarks in neutron star matter will further
reduce the maximum mass, which is being sensitive to the $\Theta^+$
potential. The pentaquark $\Theta^+$ appears around $4n_0$ for a
potential depth of $U(\Theta^+)=-100$~MeV at $n_0$. For the maximum
mass star the $\Theta^+$ population amounts to 5\% in the core. Present
confirmed pulsar mass limits, however, do provide a very weak constraint on
$\Theta^+$ potential (e.g.\ for $M>1.6 M_\odot$, the potential depth
should $U(\Theta^+) > -190$~MeV) which are a much stronger for a
hypothetical negatively charged $\Theta^-$.


\section{Hyperons and Gravitational Wave Emission}


There is an astonishing connection between microscopic reactions
involving hyperons and the overall stability of rotating neutron stars
with respect to gravitational wave emission. As pointed out by Jones
\cite{Jones:2001ie,Jones:2001ya} the {\em dominant} contribution to the
bulk viscosity of neutron star matter originates from nonmesonic weak
decay reactions of hyperons in the dense medium, not from purely
nucleonic reactions. In particular the reaction $\Lambda + p\to n+p$
\cite{Lindblom:2001hd}, and to some extent the reaction $\Sigma^- + p \to
n + n$ \cite{Haensel:2001em}, control the overall bulk viscosity
important for the r-mode instability of rotating neutron stars. The
oscillating neutron star is out of weak equilibrium and is readjusted by
those weak reactions back to equilibrium. The nonmesonic weak hyperon
decays are able to stabilise the rotating neutron star in a broader
region in the temperature-period diagram than the standard weak
processes for nucleons only. If the neutron star is unstable with
respect to the r-modes, it will emit gravitational waves. Therefore, the
knowledge on the weak nonmesonic reaction rates is crucial for
determining the stable regimes of rotating neutron stars. In addition,
two-body interactions with hyperons and the size of pairing gaps need to
be known to check for hyperon superfluidity which will substantially
change the bulk viscosity.
Improved calculations of the hyperon bulk viscosity have been performed
in \cite{vanDalen:2003uy}. Applications to hybrid stars, compact stars
with quark matter in the core, including the effects from hyperons have
been studied in \cite{Drago:2003wg}. It was demonstrated that hyperons
are very important even for the stability of hybrid stars with respect
to gravitational wave emission. In certain cases, it seems possible that
accreting rotating neutron stars persistently emit gravitational waves
\cite{Nayyar:2005th}. Even the effect of hyperon-hyperon interactions on
the r-mode stability were investigated \cite{Chatterjee:2006hy}. 

The weak nonmesonic decays $\Lambda+N\to N+N$ have been studied in
medium to heavy hypernuclei as it is the main decay channel (for reviews
see \cite{Oset:1998xz,Alberico:2001jb}). There was a long-standing
puzzle of the branching ratio of proton- and neutron-induced weak
hyperon decays, which was solved by a careful analysis of two- and
three-body processes, see \cite{Garbarino:2003yq}. The new hypernuclear
data on their weak decays indicate that the proton-induced weak decay,
which is studied for the r-modes of pulsars, is the main decay channel.
However, the neutron-induced one is nearly equally strong and is usually
neglected for the calculation of the bulk viscosity of neutron star
matter. Also, other nonmesonic processes appear in hyperon star matter
as $\Lambda + \Lambda \to \Lambda + n$ which will be determined by the
future double $\Lambda$ hypernuclear experiments. The measurement of the
weak nonmesonic decay for $S=-2$ systems is also important to test the
SU(3) symmetry of the weak matrix elements. The standard ansatz fails in
describing the weak decay amplitudes of hyperons in the vacuum and a
more general SU(3) scheme is needed \cite{SchaffnerBielich:1999sy}. It
would be interesting to test this symmetry pattern for branching ratios
of double-hypernuclei and to explore the relation to the stability of
pulsars with respect to gravitational wave emission.  

There is already some astrophysical data available on the gravitational
wave emission from pulsars. Oscillations with a frequency of 1122~Hz
have been observed for an accreting x-ray binary \cite{Kaaret:2006gr}.
If this is the rotation frequency of the neutron star, then exotic
matter must be present inside with a suitable bulk viscosity stabilising
the rotating neutron star \cite{Drago:2007iy}. The LIGO collaboration
has set new limits on the gravitational wave emission from 78 pulsars,
rotation-powered neutron stars, which are getting close to the spin-down
limit \cite{Abbott:2007ce} and which will be improved considerably in
the near future.


\section{Summary}


Hyperons have a substantial impact on neutron star properties. There is
a sizable decrease in the maximum mass of neutron stars due to the
presence of hyperons in the core. The $\Lambda$ hyperons appear at
$n\approx 2n_0$ in neutron star matter.  The population of $\Sigma$
hyperons hinges crucially on their in-medium potential. They are likely
to be absent for a repulsive potential, but the negatively charged
$\Sigma^-$ could be the first exotic component in neutron star matter
for an attractive potential.  A tiny amount of hyperons can suffice to
cool neutron stars rapidly by the hyperon direct URCA process, which is
controlled by hyperon pairing gaps. A strongly attractive YY
interaction, between $\Xi$ hyperons, results in a first order phase
transition from neutron-rich to hyperon-rich matter. This transition
allows for a new, stable solution for compact stars, hyperon stars, with
similar masses but smaller radii. The presence of the nonmesonic weak
decay reactions with hyperons in neutron stars determines the bulk
viscosity of neutron star matter and leads to an enhanced stability
window with regard to r-modes of pulsars.

It is obvious, that hypernuclear physics provides essential input for
compact star physics. The YN interactions, in particular the potential
depth in bulk nuclear matter, controls the population of hyperons for
massive neutron stars, the first exotic component likely to appear for
supranuclear densities present in the core. The emergence of hyperons
softens the nuclear equation of state and the maximum neutron star mass
possible considerably which depends on the YN coupling strength and
sensitively on the hyperon three-body forces.  Two-body YY interactions
regulate the cooling behaviour of massive neutron stars, as the hyperon
direct URCA reaction is suppressed by hyperon gaps. Nonmesonic weak
decay of hyperons in the dense medium as well as hyperon superfluidity
controls the bulk viscosity of neutron star matter which regulates the
stability of r-modes of pulsars and the emission of gravitational waves.
In addition, hyperons can generate a new class of compact stars, hyperon
stars, for a suitably attractive YY potential. The ongoing and future
experimental hypernuclear programs at DA$\Phi$NE, Jefferson Lab, KEK,
J-PARC, MAMI-C, and at GSI, Darmstadt, in particular the HypHI program
and HYPER-GAMMA with PANDA at FAIR, will provide here the decisive
inputs for addressing the macrophysics and microphysics of neutron stars
as hyperons play such an important role for many compact star
observables.

\section*{Acknowledgements}

I thank Avraham Gal for careful reading and for most valuable comments.


\bibliographystyle{h-elsevier2}
\bibliography{all,literat,hyp2006}

\end{document}